\documentclass{phb-proc4-auth}
\usepackage{graphicx}
\usepackage{amssymb}

\begin{document}
\begin{frontmatter}

%----------------------------------------------------------------------
% Specify destination and version number of the manuscript
%

\journal{SCES '04}

%----------------------------------------------------------------------
% Title of manuscript
%

\title{Charge order in the incommensurate compounds $S\!r_{14-x}C\!a_xC\!u_{24}O_{41}$}
 \author{Marie-Bernadette Lepetit\corauthref{1} }
 \corauth[1]{Corresponding Author: Marie-Bernadette Lepetit.   
              Phone:~(+33) 5.61.55.60.46. 
              Fax:~(+33) 5.61.55.60.65.  Email:~Marie@irsamc.ups-tlse.fr}
 \author{Alain Gell\'e}

 \address{Laboratoire de Physique Quantique,
              Universit\'e Paul Sabatier, 118 route de Narbonne,
              F-31062 Toulouse Cedex~4, France}

%----------------------------------------------------------------------
% Text of abstract
%

\begin{abstract}

The present paper studies, using ab-initio calculations, the influence
of the incommensurate structural modulations on the low energy physics
of the $S\!r_{14-x}C\!a_xC\!u_{24}O_{41}$ oxides.  On-site,
nearest-neighbor and next-nearest-neighbor effective parameters were
computed within a $t-J+V$ model based on the copper oxide layers. The
structural modulations appear to be the key degree of freedom,
responsible for the low energy properties such as the electron
localization, the formation of dimers in the $x=0$ compound or the
anti-ferromagnetic order in the $x=13.6$ compound.

\end{abstract}

%----------------------------------------------------------------------
% Manuscript keywords
%
% Please give two or three keywords in the form: keyword \sep keyword
% e.g. NMR \sep superconductivity
%
% NB The syntax is different from the abstract document class
%


%----------------------------------------------------------------------
% End of front page

\end{frontmatter}

%----------------------------------------------------------------------
% Manuscript text
%

The $S\!r_{14-x}C\!a_xC\!u_{24}O_{41}$ compounds are composed of
alternate layers of weakly-coupled spin ladders and spin
chains~\cite{struc}. The two sub-systems have incommensurate
translation vectors in the longitudinal chains and ladders
direction with a pseudo-periodicity is of 10 chain units for 7 ladder
units.  The misfit of  the two-subsystems
induces a structural modulation of each of them with the periodicity
of the other.  The low energy properties of the
$S\!r_{14-x}C\!a_xC\!u_{24}O_{41}$ compounds are different both from
those of uniform ladders and of uniform chains. In addition their
electronic structure is strongly dependent on the substitution of the
strontium by the isovalent calcium counter-ion.

Formal-charge analysis shows that these systems are intrinsically
doped with six holes per formula unit (f.u.).  Similar to high-$T_c$
superconductors, the spins are supported by $3d$ orbitals of the
$C\!u^{2+}$ ions while the holes are expected to be mainly supported by
the oxygen $2p$ orbitals and to form Zhang-Rice singlets~\cite{ZR}
(ZRS).  The hole repartition between the chain and ladder subsystems
is still under debate. For the undoped compound, Madelung potential
calculations~\cite{Mad97} suggest a localization of the 6 holes on
the chains. X-ray~\cite{XRay00} and optical conductivity~\cite{COpt97}
measurements however suggest that about one hole per f.u. may be
located on the ladders. Under increasing $C\!a$ substitution, an
increasing transfer of part of the holes to the ladders is
observed~\cite{XRay00,COpt97,RMN98}.  X-rays suggest that only $1.1$
hole is transferred for the $x=12$ compound, while optical
conductivity and NMR show a larger hole transfer (resp. $2.8$ for
$x=11$ and $3.5$ for $x=11.5$).

The chain subsystem is particularly intriguing since not only it
exhibits a spin gap ($\Delta_\sigma\simeq 11-12\
meV$~\cite{Magn96B,RMN97} for $x=0$), but it also presents a low
temperature magnetic behavior strongly dependent of the $C\!a$
content. Indeed, the undoped compound exhibit a charge order at $T<
240K$~\cite{Thermo00} as well as spin
dimerization~\cite{Magn96B}. These dimers are seen to be
ordered~\cite{RMN98B,Neut98} according to a 5 sites pseudo-periodicity
where second-neighbor dimers are separated by two ZRS. The charge
order disappear with $C\!a$ doping ($x>8$) even-though the magnetic
interactions within and between the dimers seem to remain
unchanged. At large doping ($x \ge 11$) and very low temperatures
($<2.5K$) an antiferromagnetically ordered phase is observed~\cite{Magn99}.

These properties cannot be accounted for in a homogeneous hole-doped
spin-chain model. One can thus look for their origin toward the
effects of the incommensurate structural modulations. We thus evaluated
the parameters of a second neighbor $t-J+V$ model, as a function of
the modulation parameter, using embedded-fragment spectroscopy ab
initio calculations~\cite{DDCI}. The calculations have been
performed for two different doping, namely for the
$S\!r_{14}C\!u_{24}O_{41}$ and $S\!r_{0.4}C\!a_{13.6}C\!u_{24}O_{41}$
compounds in their low temperature phase~\cite{struc}.

As expected the NN effective exchange are essentially ferromagnetic in
nature while the second neighbor exchanges are antiferromagnetic and
weakly modulated.  In the $x=0$ compound, the NN exchange modulations 
are weak (standard deviation $\sim 12\%$ of the average value: $21.3\
meV$). On the contrary, in the $x=13.6$ system the modulations are
very large (standard deviation $\sim 110\%$ of the average value: $10\
meV$). For a small part of the modulation distribution the distortions
are such that the NN exchange becomes antiferromagnetic. Similarly the
hopping integrals are very strongly modulated.

\begin{figure}[h] %%%%% Analyse de Fourier des energies orbitalaires
\centering
\resizebox{!}{2.5cm}{\includegraphics{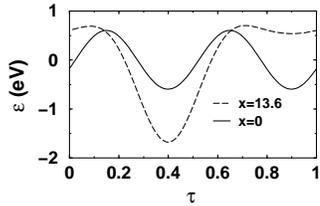}}
\caption{Orbital energies as a function of the structural
 chain modulation parameter $\tau$.}
\label{f:e}
\end{figure}

The main surprise however is the modulations of the on-site orbital
energies (see figure~\ref{f:e}). Indeed, they vary in a range of a few
electron-Volts, that is much larger than any other parameter in the
model (the next largest effective integrals are the NN hopping
integrals that range between $60\ meV$ and $200\ meV$ for $x=0$
and from $-70\ meV$ to $250\ meV$ for $x=13.6$). This strong
on-site energy variation is thus dominating the compounds physics
through the spin localization on the low energy sites.

\begin{figure} %% remplissages
\resizebox{8cm}{3cm}{\includegraphics{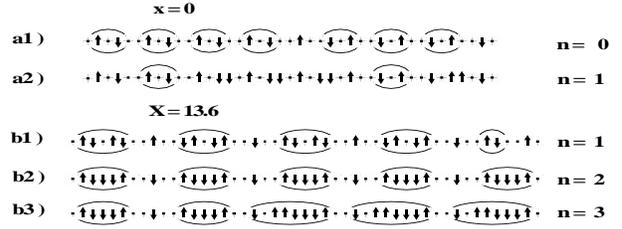}}
%\vspace*{0.5eM}
\caption{Localization of the magnetic electrons along the chain, as a
function of the number $n$ of holes transfered to the ladders. a)
for $x=0$, b) for $x=13.6$. Squares stand for ZRS, the ellipsoids
delimit magnetically stable spin clusters.}
\label{rempl}
\end{figure}

Let now use these energies to locate the spins along the chains for
several hypotheses of hole filling, namely $n=0,1$ per f.u. for
$x=0$ and $n=1,2,3$ for $x=13.6$ (see figure~\ref{rempl}).
For $x=0$ and all the holes on the chains, the computed spin
arrangement retrieves the experimentally observed second-neighbor
dimeric units as well as the correct number of free spins (computed~:
$0.5$ per f.u., magnetic susceptibility measurements~: $0.55$). For
$x=13.6$, one observes the formation of low-spin clusters, with NN
spins and no second-neighbor dimeric units. It it noticeable that the
existence of antiferromagnetic NN exchanges allow these clusters to be
low spin and to present low frustration of both NN and NNN exchanges
(at most one frustrated NNN interaction).  These results should be put
into perspective with the antiferromagnetic ordering seen in magnetic
susceptibility and ESR measurements. For $n=1,2$ one still observes a
large number of free spins, while for $n=3$ they essentially
disappeared, in better agreement with  magnetic susceptibility
experiments~\cite{Magn99}. 

We have studied the importance of the structural modulations on the
low energy physics of the $S\!r_{14-x}C\!a_{x}C\!u_{24}O_{41}$
family. Surprisingly these distortions are not simply responsible for
weak parameters modulations around their average value, but induce
very large variations of the orbital energies, effective hoppings and
exchanges. This is the variation of the orbital energies that is
responsible for the low energy properties of the compounds, through
the localization of the magnetic electrons. It is in particular
responsible for the formation of second neighbor dimers in the undoped
compound and of stable low-spin clusters in the highly doped one.

%----------------------------------------------------------------------
% Reference section
%

\end{document}